\documentclass[reqno]{amsart}%
\usepackage{amsfonts}
\usepackage{verbatim}
\usepackage{xcolor}
\usepackage{amsmath}
\usepackage{amssymb}
\usepackage{graphicx}
\usepackage{accents}%
\providecommand{\U}[1]{\protect\rule{.1in}{.1in}}
\newtheorem{theorem}{Theorem}[section]

\newtheorem{corollary}[theorem]{Corollary}
\newtheorem{definition}[theorem]{Definition}

\numberwithin{equation}{section}
\begin{document}
\title[KdV soliton gas]{Continuous binary Darboux transformation as an abstract framework for KdV
soliton gases\\ }
\dedicatory{This work is dedicated to Lili (Olivier) Kimmoun and Vladimir Zakharov, both
of whom passed away in 2023. Their deep and lasting influence on nonlinear
water waves and dispersive dynamics continues to shape the field and inspire
future research.}\author{Alexei Rybkin}
\address{Department of Mathematics and Statistics, University of Alaska Fairbanks, PO
Box 756660, Fairbanks, AK 99775, USA}
\email{arybkin@alaska.edu}
\thanks{The author is supported in part by the NSF grant DMS-2307774. The author would
like to thank the Isaac Newton Institute for Mathematical Sciences for support
and hospitality during the programme Dispersive Hydrodynamics when this work
was originated (EPSRC Grant Number EP/R014604/1).}
\subjclass{34L25, 37K15, 47B35}
\keywords{Soliton gas, KdV equation, Darboux transformation.}
\date{December 2025}

\begin{abstract}
We present a unified operator-theoretic framework for constructing
deterministic KdV soliton gases and step-type KdV solutions. Starting from
Dyson's determinantal formula, we obtain a broad class of reflectionless
solutions and describe their basic spectral and analytic properties, including
their interpretation as deterministic soliton gases. We then introduce a
continuous binary Darboux transformation that acts directly on the scattering
data and generates general step-type solutions, with particular emphasis on
reflectionless hydraulic-jump-type profiles modelling a soliton condensate on
the left and vacuum on the right. The paper is methodological in nature: our
goal is not to develop a full kinetic or probabilistic theory, but to show how
classical tools from spectral and scattering theory can be combined into a
conceptually simple framework that accommodates both reflectionless and
non-reflectionless soliton gas configurations, including step-like backgrounds.

\end{abstract}
\maketitle

\section{\medskip Introduction}

The concept of a \emph{soliton gas} originates in the pioneering work of
Zakharov and collaborators in the early 1970s, where the idea of interpreting
large ensembles of solitons as a macroscopic statistical medium was first
articulated; see Zakharov~\cite{Zakharov(soliton gas)71} and the
monograph~\cite{NPZ}. In the KdV setting, solitons correspond to simple
negative eigenvalues of the one-dimensional Schr\"odinger operator
\[
\mathbb{L}_{q}=-\partial_{x}^{2}+q(x),\qquad-\infty<x<+\infty,
\]
and a soliton gas is understood as the thermodynamic limit of an ensemble of
such eigenvalues. A defining feature of this picture is that the collective
behavior of the soliton ensemble can be described by a nonlinear kinetic
equation, whose derivation relies on pairwise phase shifts and weak spectral
inhomogeneity. This viewpoint, developed in particular by
El--Kamchatnov~\cite{ElKom05} in the mid-2000s and further refined by El with
collaborators in \cite{El2016,El020,El21}, has placed soliton gases at the
center of modern studies of integrable turbulence and dispersive
hydrodynamics; see also the rigorous work of Girotti--Grava--Jenkins--
McLaughlin~\cite{Grava21} and the experimental results of Costa--Osborne et
al.~\cite{CostaOsbornePRL14} and Redor et al.~\cite{RedoretalPRL19}. For a
recent survey we refer to Suret et al.~\cite{Surretetal2024}.

From the physical point of view, the KdV equation is a universal model for the
unidirectional propagation of long, weakly nonlinear and weakly dispersive
water waves in shallow water regimes. Solitons describe coherent structures in
surface and internal waves, nearshore hydrodynamics, and tsunami propagation.
In this context, a soliton gas provides a statistical description of random
water-wave fields in terms of interacting soliton components and thus forms a
central element of the emerging theory of integrable turbulence.

Historically, nearly all early constructions of soliton gases assume that the
background potential is zero: solitons propagate on a zero background. The
asymptotic behavior at $\pm\infty$ is then identical, so the scattering
problem is \emph{symmetric} in the following sense: both spatial infinities
share the same zero background, the absolutely continuous spectrum is the
single interval $[0,\infty)$, and solitons correspond to isolated negative
eigenvalues. The inverse scattering transform (IST), originating in the work
of Gardner--Greene--Kruskal--Miura~\cite{GGKM67} and Zakharov--Shabat and
systematically developed in monographs such as~\cite{AC91,NPZ}, admits a
well-controlled thermodynamic limit in this setting, and the kinetic
description follows from the Marchenko theory and its Riemann--Hilbert
refinements (see, e.g.,~\cite{GT09}).

In many physically relevant situations, however, the potential does not decay
but instead exhibits a \emph{step-like} structure,
\[
q(x,t)\rightarrow c_{-}\quad(x\rightarrow-\infty),\qquad q(x,t)\rightarrow
c_{+}\quad(x\rightarrow+\infty),
\]
with $c_{-}\neq c_{+}$. In this case, the scattering problem becomes
\emph{asymmetric} in the sense that the limiting backgrounds at $\pm\infty$
differ. The analytic theory of KdV with step-like initial conditions,
originating in Khruslov 1976~\cite{Hruslov76} and further developed in more
recent works such as Egorova--Michor--Teschl \cite{Egorova2022} and
Ablowitz--Luo--Cole~\cite{AblowitzLuoCole2018}, reveals several spectral
features absent in the symmetric (decaying) case:

\begin{itemize}
\item the continuous spectrum is the union of two shifted half-lines
$[-h_{-}^{2},\infty)\cup\lbrack-h_{+}^{2},\infty)$ with $h_{\pm}$ explicitly
computable in terms of $c_{\pm}$;

\item the Jost solutions, reflection coefficients, and transmission
coefficients can be properly defined but become asymmetric and require
independent left and right scattering formulations;

\item depending on the sign of $c_{+}-c_{-}$, solitons may propagate only on
one side or become trapped by the band edge of the continuous spectrum;

\item the discrete spectrum interacts nontrivially with the band edges,
necessitating a reconsideration of how spectral densities should be normalized.
\end{itemize}

These features show that the conventional definition of a soliton gas cannot
be transferred verbatim to the step-like setting. Recent work on finite-gap
thermodynamic limits and soliton condensates (for example,
El--Taranenko~\cite{ElTaranenko2020} and Congy--El--Roberti--Tovbis
\cite{Condi22}) suggests that a meaningful generalization should involve a
two-component kinetic structure reflecting the two asymptotic backgrounds. See
also the recent paper by Bertola--Jenkins--Tovbis \cite{Bertoleetal23} in this context.

In this paper we take a different perspective. Our aim is not to construct a
full statistical theory or a kinetic equation for soliton gases on step-like
backgrounds. Instead, this short note is of \emph{methodological} character:
we show how existing analytic structures, in particular the continuous binary
Darboux transformation recently put forward in \cite{RybCBTD23}, may be used
to organize \emph{deterministic} soliton gas ensembles in the presence of
left--right asymmetry. Recall that the deterministic notion of soliton gas
goes back to Zakharov's 1971 introduction of a continuous spectral density
describing an infinite ensemble of KdV solitons. More recently, Zakharov and
collaborators (see, e.g., \cite{ZakharovetalPhysD2016}) put forward ways to
generate deterministic soliton gases by means of structured superpositions of
dressing operations, which are now viewed as a natural complement to
statistical soliton gas theory and play a role in the rigorous formulation of
integrable turbulence.

Our main observation is that the continuous binary Darboux transformation (its
discrete counterpart goes back to classical work of Babich--Matveev--Salle and
the general theory of Darboux
transformations~\cite{Babichetal85,MatveevSalle91,GuetalBook05}) admits a
formulation compatible with step-like scattering theory and captures the
interaction of soliton ensembles with both asymptotic backgrounds. This provides:

\begin{itemize}
\item a structural description of how elementary dressing operations compose
under asymmetric scattering data;

\item a natural way to define deterministic soliton gas densities that respect
the left--right decomposition of the spectrum;

\item a unifying viewpoint that includes the classical decaying case, the
step-like case, and the emerging finite-gap and condensate regimes.
\end{itemize}

Thus the contribution of this note is methodological: it identifies a
conceptually clean framework in which deterministic soliton gases may be
constructed and in which possible extensions toward statistical and kinetic
descriptions can be organized.

\bigskip In what follows we develop this framework in four steps. In
Section~\ref{sec:Dyson} we recall Dyson's construction and its relation to
classical multi-soliton solutions. Section~\ref{sec:Refl} interprets these
solutions as reflectionless potentials and records some analytic consequences.
Section~\ref{sec:Gas} introduces deterministic soliton gases in this setting
and discusses reflectionless step-like potentials and their interpretation as
condensate--vacuum configurations related to undular bores. Finally,
Section~\ref{sec:Step} extends the framework to general step-type potentials
via a continuous binary Darboux transformation and formulates several open
problems. We conclude with a brief discussion of how this approach fits into
the broader soliton gas and integrable turbulence literature.

\section{Dyson formula}

\label{sec:Dyson}

In this section we recall Dyson's determinantal formula for constructing KdV
solutions from a nonnegative measure on the positive half-line. We also
explain how the formula relates to classical soliton solutions and earlier
approaches of Bargmann, Lundina, and Marchenko. The point of view is that
Dyson's construction already provides a natural deterministic soliton gas
associated with a given spectral measure.

Let $\sigma(k)$ be a compactly supported nonnegative measure on $[0,\infty)$
such that
\[
\mathrm{d}\sigma(k)\geq0,\qquad\int_{0}^{\infty}\frac{\mathrm{d}\sigma(k)}%
{k}<\infty.
\]
Such measures are also known as Carleson measures (see, e.g., \cite{Koosis}).
For each time $t\in\mathbb{R}$ introduce
\[
\mathrm{d}\sigma_{t}(k)=\exp(8k^{3}t)\,\mathrm{d}\sigma(k),
\]
and define a two-parameter $(x,t)$ integral operator $\mathbb{K}_{x,t}$ on
$L^{2}(\mathrm{d}\sigma_{t})$ by
\[
(\mathbb{K}_{x,t}f)(s)=\int_{0}^{\infty}\frac{e^{-(s+k)x}}{s+k}%
\,f(k)\,\mathrm{d}\sigma_{t}(k),\qquad f\in L^{2}(\mathrm{d}\sigma_{t}).
\]
The operator $\mathbb{K}_{x,t}$ is Hankel, and its significance is expressed
by the following fundamental result (which is a particular case of Theorem
\ref{MainThm} below).

\begin{theorem}
[Dyson formula]\label{Thm: Dyson} The operator $\mathbb{K}_{x,t}$ is trace
class, and the function
\begin{equation}
q_{\sigma}(x,t)=-2\partial_{x}^{2}\log\det(I+\mathbb{K}_{x,t})\text{ (Dyson's
formula)} \label{Dyson}%
\end{equation}
is a classical solution to the KdV equation
\begin{equation}
\partial_{t}u-6u\partial_{x}u+\partial_{x}^{3}u=0,\qquad x,t\in\mathbb{R}.
\label{KdV}%
\end{equation}

\end{theorem}

We record several historical remarks and connections.

\begin{itemize}
\item If $\sigma$ is a finite sum of Dirac masses,
\[
\mathrm{d}\sigma(k)=\sum_{1\leq n\leq N}c_{n}^{2}\,\delta(k-\kappa
_{n})\,\mathrm{d}k,\qquad\kappa_{n}>0,\ c_{n}>0,
\]
then $\mathbb{K}_{x,t}$ becomes an $N\times N$ matrix $K_{x,t}$ with entries
\[
K_{x,t}(n,m)=\frac{c_{n}c_{m}}{\kappa_{n}+\kappa_{m}}e^{-(\kappa_{n}%
+\kappa_{m})x+4(\kappa_{n}^{3}+\kappa_{m}^{3})t},
\]
and (\ref{Dyson}) reduces to the classical Kay--Moses formula for pure
$N$-soliton solutions:
\begin{equation}
q_{N}(x,t)=-2\partial_{x}^{2}\log\det(I+K_{x,t}). \label{Kay Moses formula}%
\end{equation}

\item If $\sigma$ is discrete but infinite (i.e. $(\kappa_{n})\in\ell^{\infty
}$, $\sum c_{n}^{2}/\kappa_{n}<\infty$), then (\ref{Dyson}) recovers the 1992
Gesztesy--Karwowski--Zhao construction \cite{Gesztesy-Duke92} based on certain
limiting procedures for (\ref{Kay Moses formula}).

\item For a specific absolutely continuous measure $\sigma$, in our form
(\ref{Dyson}), Dyson's formula was used in 1986 by Venakides \cite{Venak86}
(where it is referred to as the Bargmann formula) with reference to previous
works. Dyson's famous 1976 paper \cite{Dyson1976} however is not mentioned
therein. We refer to (\ref{Dyson}) as Dyson's formula as it is also well known
in the context of random matrices.

\item The substitution
\[
q(x,t)=-2\partial_{x}^{2}\log\tau(x,t),
\]
where $\tau$ is the \emph{Hirota tau-function}, is classical in the theory of
integrable systems. A family of finite-gap KdV solutions was also expressed in
the same form in the seminal 1975 Its--Matveev paper \cite{ItsMatveev1975},
where the tau function $\tau(x,t)$ is expressed in terms of the Riemann theta
function associated with an underlying hyperelliptic Riemann surface.

\item If we drop the condition $\mathrm{d}\sigma(k)\geq0$, Dyson's formula may
still produce a solution, but $q_{\sigma}$ becomes singular. For example, if
$\mathrm{d}\sigma(k)=-\delta(k-1/2)\,\mathrm{d}k$, then
\[
q_{\sigma}(x,t)=-\partial_{x}^{2}\log(1-e^{t-x})^{2},
\]
which has a moving real double pole at $x=t$. Thus the method offers a
convenient way to study singular KdV solutions (see, e.g., Ma 2005 \cite{Ma05}).
\end{itemize}

What is important for our purposes is that Dyson's formula provides a natural
deterministic ``soliton gas'' construction (see Section \ref{sec:Gas}): the
measure $\sigma$ selects an ensemble of pure KdV soliton solutions, and
(\ref{Dyson}) describes the resulting superposition in a form consistent with
the integrable structure.

\section{Reflectionless solutions}

\label{sec:Refl}

In this section we interpret the solutions produced by Dyson's formula as
reflectionless potentials in the sense of modern spectral theory. We review
known generalizations, state their analytic properties, and note uniqueness
and regularity consequences relevant for soliton gases. This perspective will
motivate our definition of deterministic soliton gases in the next section.

Historically, a reflectionless potential is a potential in a full-line
Schr\"odinger scattering problem whose reflection coefficient vanishes
identically on the continuous spectrum.

This notion has been extended beyond classical scattering in the work of
Lundina \cite{Lundina85} and Marchenko \cite{Marchenko91}, where such
potentials are called \emph{generalized reflectionless} and are described
using the Titchmarsh--Weyl $m$-function. Formula (\ref{Dyson}) produces a
notion of generalized reflectionless potentials reminiscent of the
constructions due to Lundina \cite{Lundina85} and Marchenko \cite{Marchenko91}%
. In particular, \cite{Marchenko91} shows that if the integral equation
\begin{align}
&  e^{-4\kappa^{3}t+\kappa x}\left\{  a(\kappa)y(\kappa)-\frac{1}{2\kappa
}\left[  \int\frac{y(s)-y(\kappa)}{s-\kappa}\,\mathrm{d}\sigma(s)-1\right]
\right\} \label{MarhLundina formula}\\
&  =e^{4\kappa^{3}t-\kappa x}\left\{  [a(\kappa)-1]\,y(-\kappa)-\frac
{1}{2\kappa}\left[  \int\frac{y(s)-y(-\kappa)}{s+\kappa}\,\mathrm{d}%
\sigma(s)-1\right]  \right\}  ,\nonumber
\end{align}
is uniquely solvable for $y(\kappa,x,t)$, then
\begin{equation}
q(x,t)=-2\partial_{x}\int y(\kappa,x,t)\,\mathrm{d}\sigma(\kappa) \label{q}%
\end{equation}
satisfies the KdV equation with $q(x,0)=q(x)$, where $a$ and $\sigma$ encode
the scattering data of $q(x)$. The relation between (\ref{MarhLundina formula}%
), (\ref{q}) and Dyson's formula (\ref{Dyson}) is not evident (at least to us)
and is worth investigating, especially in view of an open question concerning
$\sigma$ posed in \cite{Marchenko91}. We emphasize that the solvability of
(\ref{MarhLundina formula}) is far from trivial. Finally, the methods of
\cite{Marchenko91} require smoothness of $q(x)$, which is not needed in the
present framework.

Further generalizations (also based on the $m$-function) include
reflectionless potentials on sets smaller than $(0,\infty)$, for example on
band spectra. See Hur--McBride--Remling~\cite{HurMcBrideRemling2016} for a
rigorous treatment and Johnson--Zampogni~\cite{JohnsonZampogni2015} for an
extensive bibliography. We also refer to Kotani \cite{Kotani2008}, which is
closest in spirit to our considerations.

The following statement follows from \cite{RybCBTD23}.

\begin{theorem}
[Reflectionless potentials]\label{Thm: refl} Let $q_{\sigma}(x,t)$ be as in
Theorem \ref{Thm: Dyson}. Then the full-line Schr\"{o}dinger operator
\[
\mathbb{L}_{q_{\sigma}}=-\partial_{x}^{2}+q_{\sigma}(x,t)
\]
is reflectionless on $(0,\infty)$, and its spectrum is
\[
\operatorname{Spec}(\mathbb{L}_{q_{\sigma}})=\{-k^{2}:k\in\operatorname{Supp}%
(\sigma)\}\cup\lbrack0,\infty).
\]

\end{theorem}

\bigskip

Since our $q_{\sigma}$ is obtained as a uniform limit of pure soliton
potentials, the following statements hold.

\begin{corollary}
[Analyticity]If $h=\sup\operatorname{Supp}(\sigma)>0$, then $q_{\sigma}(z,t)$
is real analytic in the strip $|\operatorname{Im}z|<1/h$ and satisfies the
universal bounds
\[
|q_{\sigma}(x+iy,t)|\leq2h^{2}(1-h|y|)^{-2},\qquad-2h^{2}<q_{\sigma}(x,t)<0.
\]

\end{corollary}

Note that $q_{\sigma}(x,t)$ need not decay (or even have a limit) as
$x\rightarrow-\infty$ and therefore typically lies outside the classical
scattering framework (but still within an asymmetric scattering setting as
discussed above).

\begin{corollary}
[Lundina 1985]For fixed $h>0$, the family of analytic functions
\[
\{\,q_{\sigma}:\operatorname{Supp}(\sigma)\subseteq\lbrack0,h]\,\}
\]
is normal (that is, locally uniformly precompact).
\end{corollary}

\begin{corollary}
If $\inf\operatorname{Supp}(\sigma)>0$, then $q_{\sigma}(x,t)$ decays
exponentially as $x\rightarrow+\infty$.
\end{corollary}

\begin{corollary}
[Uniqueness]Reflectionless solutions to KdV are unique.
\end{corollary}

Note that the last corollary is a highly nontrivial statement. As was shown by
Cohen--Kappeler in their 1989 paper \cite{Cohen1989}, rapid decay of initial
data at $+\infty$ and smoothness does not guarantee uniqueness. It was proved
in the recent work of Chapouto--Killip--Visan \cite{ChapoutoKillipVisan2024}
that smoothness (even continuity) and boundedness imply uniqueness, which
holds in our case.

The properties stated in these corollaries translate into structural
properties of deterministic soliton gases, which we discuss next.

\section{Deterministic KdV soliton gas}

\label{sec:Gas}

This section explains how the reflectionless solutions generated by Dyson's
formula give rise to deterministic soliton gases for KdV. We summarize their
spectral character and analytic structure and relate the construction to the
primitive-potential framework of Zakharov. The emphasis is on how the spectral
measure $\sigma$ encodes the macroscopic soliton distribution.

Following modern terminology (see El--Taranenko (2020) \cite{ElTaranenko2020}%
), a \emph{deterministic KdV soliton gas} is a reflectionless KdV solution
whose negative spectrum contains a continuous interval $\left[  -a^{2}%
,-b^{2}\right] $, $b>0$, with a prescribed spectral density. This notion goes
back to Zakharov's 1971 paper \cite{Zakharov(soliton gas)71}, where the
soliton distribution function was first introduced.

The gas is called ``deterministic'' because the soliton spectrum is described
by a macroscopic spectral density rather than stochastic eigenvalue statistics.

We adopt a broader viewpoint and call any bounded reflectionless KdV solution
a deterministic soliton gas.

\begin{definition}
[Deterministic soliton gas]We call a KdV soliton gas deterministic if it is
generated by Dyson's formula (\ref{Dyson}).
\end{definition}

Observe that if $\operatorname{Supp}(\sigma)$ is a finite union of disjoint
closed intervals and $0\notin\operatorname{Supp}(\sigma)$, then $q_{\sigma}$
is a deterministic soliton gas. Indeed, $\int\mathrm{d}\sigma(k)/k<\infty$
holds automatically.

Also note that in Zakharov's terminology (see, e.g., Nabelek--Zakharov (2016)
\cite{ZakharovetalPhysD2016}), $q_{\sigma}$ corresponds to a primitive
potential with $R_{2}=0$, one of the two dressing functions. The case
$R_{2}\neq0$ remains challenging, although it is tractable in the symmetric
setting $R_{1}=R_{2}$ \cite{NabelikZakharov2019}.

\subsection*{Properties of deterministic soliton gases}

The considerations of the previous section immediately imply several general
properties of deterministic soliton gases.

\begin{itemize}
\item As an analytic function, $q_{\sigma}(x,t)$ is completely determined by
its values on any subset of positive Lebesgue measure.

\item By uniqueness of reflectionless solutions, a deterministic soliton gas
never bifurcates.

\item We have $-2h^{2}<q_{\sigma}(x,t)<0$, so solitons do not pile up.

\item Since
\[
\operatorname{Spec}(\mathbb{L}_{q_{\sigma}})=\{-k^{2}:k\in\operatorname{Supp}%
(\sigma)\}\cup\lbrack0,\infty),
\]
the solution $q_{\sigma}(x,t)$ decays as $x\rightarrow+\infty$ (see, for
example, Remling \cite{Remling2008}).
\end{itemize}

Thus, loosely speaking, a deterministic soliton gas is deterministic in two
senses: its spectral density is prescribed, and the resulting solution is
completely determined by any nontrivial fragment of its spatial profile. These
and other structural properties of deterministic soliton gases are rarely
discussed explicitly in the literature (at least we have not seen such discussions).

\subsection{\textbf{Reflectionless step-like potentials}}

Such potentials are particularly relevant to the study of \emph{soliton gas
condensates}. Recall that a soliton gas condensate is a maximally dense
soliton gas whose spectral density attains the upper bound allowed by the
reflectionless condition \cite{ElTaranenko2020}, \cite{Grava21}. Consider
\[
\mathrm{d}\sigma(k)=2(k/h)\sqrt{h^{2}-k^{2}}\,\mathrm{d}k,\qquad0\leq k\leq
h.
\]
Clearly, $\mathrm{d}\sigma\geq0$ and $\int_{0}^{h}\mathrm{d}\sigma
(k)/k<\infty$. Thus Theorem~\ref{Thm: Dyson} applies. The resulting
$q_{\sigma}$ may be computed either by (\ref{Dyson}) or alternatively by
\begin{align}
q_{\sigma}(x,t)  &  =8\Bigl[\int_{0}^{h}(s/h)\sqrt{h^{2}-s^{2}}e^{-2sx}%
Y(s;x,t)\,\mathrm{d}s\Bigr]^{2}\label{q ref}\\
&  \quad-8\int_{0}^{h}(s/h)^{2}\sqrt{1-s^{2}}e^{-2sx}Y(s;x,t)\,\mathrm{d}%
s,\nonumber
\end{align}
where $Y$ solves the Fredholm equation
\begin{equation}
Y(\alpha;x,t)+\int_{0}^{h}2(s/h)\sqrt{h^{2}-s^{2}}\frac{e^{8s^{3}t-2sx}%
}{s+\alpha}Y(s;x,t)\,\mathrm{d}s=1,\qquad\alpha\in\lbrack0,h]. \label{Y}%
\end{equation}
In \cite{RybCBTD23} we show that
\[
q_{\sigma}(x,t)\rightarrow-h^{2}\ \text{as }x\rightarrow-\infty,\qquad
q_{\sigma}(x,t)\rightarrow0\ \text{as }x\rightarrow+\infty.
\]
Thus $q_{\sigma}$ can be viewed as a smooth reflectionless deformation of the
``hydraulic jump'' potential
\[
q(x)=-h^{2},\ x<0;\qquad q(x)=0,\ x\geq0,
\]
a short-range perturbation of a pure step function. Its spectrum is purely
absolutely continuous,
\[
\operatorname{Spec}(\mathbb{L}_{q})=[-h^{2},\infty),
\]
with $(-h^{2},0)$ simple and $(0,\infty)$ double. Thus our $q_{\sigma}$ is a
\emph{reflectionless step-like potential}. To the best of our knowledge, this
construction did not explicitly appear in the literature prior to
\cite{RybCBTD23}.

The fact that the spectral measure appearing in our step-like reflectionless
solution agrees exactly with the density of states of a one-gap finite-gap
potential follows directly from the classical spectral theory of periodic and
quasi-periodic KdV potentials. In the oscillatory region generated by the
dispersive resolution of a step, the solution is known, beginning with the
work of Gurevich and Pitaevskii, to approach, on the fast spatial scale, a
slowly modulated cnoidal wave whose parameters evolve according to the Whitham
equations \cite{Gurevich74}. Such solutions are precisely the genus-one
finite-gap potentials described in the finite-gap/IST theory of Novikov,
Dubrovin, Matveev, Its, and Kotlyarov \cite{NPZ}, \cite{ItsKotlyarov}.

For any one-gap potential, the associated Schr\"{o}dinger operator has a
single finite spectral band, and the density of states is a universal
algebraic function determined solely by the endpoints of this band; the
geometry of the underlying hyperelliptic Riemann surface leaves no additional freedom.

Because the reflectionless step-like initial data produce, in the long-time
limit, a potential that is locally indistinguishable from such a one-gap
configuration, the corresponding local spectral problem must inherit the same
Riemann-surface structure and therefore the same density of states. In other
words, once the band edges appearing in the Gurevich--Pitaevskii modulation
are fixed, the finite-gap spectral theory forces a unique density-of-states
measure, and this is precisely the measure that arises from the thermodynamic
description of the solution. This observation is fully consistent with the
modern interpretation of dispersive shocks and their spectral structure in
terms of finite-gap theory and soliton condensates developed by El,
Kamchatnov, Tovbis, and coauthors \cite{ElKom05}, \cite{El020}.

We also note that it was proved by Khruslov (Hruslov) in 1976 that a step-like
potential produces an infinite sequence of asymptotic solitons of height
$-2h^{2}$, that is, twice the height of the initial jump. This result was
reproduced by Venakides in 1986 in \cite{Venak86}, and his arguments are based
on (\ref{Dyson}), which indicates that this phenomenon is far more general:
the fastest soliton always propagates with asymptotic velocity $2h^{2}$, where
$h^{2}=-\inf\operatorname{Spec}\mathbb{L}_{q}$. Determining the associated
asymptotic phases is considerably more delicate (work in progress).

\subsection*{Informal remarks}

The reflectionless step-like potential considered in this work provides a
particularly transparent example of a soliton condensate adjoining a vacuum
state, and its long-time evolution is the classical setting for the emergence
of an undular bore in the sense of Gurevich and Pitaevskii. On the left, the
initial data support a densely populated soliton component whose evolution
leads, inside the expanding dispersive-shock region, to the formation of a
nonlinear wavetrain locally indistinguishable from a one-gap finite-gap
solution. In this regime the soliton population reaches its maximal spectral
density, so that the field behaves as a saturated soliton condensate: the
local structure is that of a cnoidal wave whose parameters evolve smoothly
according to the Whitham modulation equations. The periodic wave forms the
interior of the undular bore, representing the fully condensed limit of a
soliton ensemble.

In contrast, the right side of the step contains no solitonic spectral
content, and thus evolves into a vacuum state with zero density. The undular
bore that develops between these two regions acts as a sharply defined
interface separating the condensate from the vacuum. Its inner region consists
of nearly harmonic oscillations transitioning continuously into a nonlinear
periodic wave of finite amplitude, while its outer region resolves into a
sequence of increasingly separated solitary pulses at the trailing edge. The
overall structure is fully described by the self-similar Gurevich--Pitaevskii
modulation solution, which enforces smooth matching of the periodic finite-gap
interior to the constant outer states. In the spectral language of integrable
systems, the bore represents an expanding region in which the system selects
the unique one-gap Riemann surface compatible with the left condensate and
right vacuum, and populates its spectral band at full capacity. Thus the
reflectionless step-like profile provides a natural and analytically tractable
model of a condensate--vacuum system, with the undular bore serving as the
dynamical mechanism through which the two phases connect.

Our measure $\sigma$ can be purely singular continuous. It would be
interesting to ask whether such a situation could have any soliton gas meaning.

A key feature of our approach is that it applies equally well to
non-reflectionless potentials. We turn to this in the next section.

\section{Step-type potentials and the continuous binary Darboux
transformation}

\label{sec:Step}

In this section we extend the Dyson construction to step-type KdV solutions by
introducing a continuous binary Darboux transformation acting on the
scattering data. This provides a mechanism for modifying (and even
redesigning) the negative spectrum while preserving the right reflection
coefficient. In this way, one can superimpose a deterministic soliton gas on a
general step-type background in a controlled manner.

We call a locally summable real function $q(x)$ a (right) \emph{step-type
potential} if

\begin{itemize}
\item its spectrum is bounded below
\begin{equation}
\inf\operatorname{Spec}(\mathbb{L}_{q})\geq-h^{2},
\label{eq boundedness below}%
\end{equation}
for some finite $h$;

\item $q(x)$ decays sufficiently fast as $x\rightarrow+\infty$ (see below).
\end{itemize}

Step-like potentials considered in the previous section are clearly step-type.
The main feature of step-type potentials is that they admit asymmetric
scattering theory: they support right Jost solutions $\psi$, i.e. for each
$\operatorname{Im}k\geq0$,
\[
\psi(x,k)\sim e^{\mathrm{i}kx},\qquad x\rightarrow+\infty,
\]
and the right reflection coefficient $R(k)$ is well defined. It is proved in
\cite{GryRybBLMS20} that

\begin{theorem}
[Grudsky--Rybkin, 2020]If
\[
\int^{\infty}x^{5/2}|q(x)|\,\mathrm{d}x<\infty\qquad\text{\emph{(faster decay
at $+\infty$)}}
\]
and if $q_{n}(x)=q(x)\big|_{(-n,\infty)}$, then
\[
q_{n}(x,t)\longrightarrow q(x,t)
\]
uniformly on compact subsets of $\mathbb{R}\times\mathbb{R}_{+}$, where
$q(x,t)$ is a classical solution to KdV. Moreover,
\[
S_{q}(t)=\Bigl\{\,R(k)e^{8\mathrm{i}k^{3}t},\;e^{8k^{3}t}\,\mathrm{d}%
\rho(k):k\geq0\Bigr\}
\]
is the scattering data for $q(x,t)$.
\end{theorem}

We call $q(x,t)$ a \emph{step-type KdV solution} with data $S_{q}
=\{R,\mathrm{d}\rho\}$. The main feature of this data is that $R(k)$ is
essentially an arbitrary function such that $R(-k)=\overline{R(k)}$ and
$\left\vert R(k)\right\vert \leq1$, while the measure $\rho$ is nonnegative
and finite and otherwise arbitrary. The time evolution of $S_{q}$ under the
KdV flow is nevertheless the same as in the classical decaying case.

Note that step-type KdV solutions decay at $+\infty$ but are essentially
arbitrary at $-\infty$. The most nontrivial fact is that such solutions never
become singular (see \cite{GruRybSIMA15}, \cite{GryRybBLMS20}).

In the context of the present paper, step-type solutions are important due to
the following statement (\cite{RybCBTD23}).

\begin{theorem}
[Continuous binary Darboux transformation]\label{MainThm} Assume that $q(x,t)$
is a step-type KdV solution with scattering data $S_{q}=\{R,\mathrm{d}\rho\}$.
Let $\sigma(k)$ be a finite signed compactly supported measure on $[0,\infty)$
satisfying
\[
\int\frac{|\mathrm{d}\sigma(k)|}{k}<\infty,\qquad\mathrm{d}\rho+\mathrm{d}%
\sigma\geq0.
\]
Define the integral operator $\mathbb{K}_{x,t}$ on $L^{2}(\mathrm{d}\sigma
_{t})$ by
\[
K_{x,t}(\lambda,\mu)=\int_{x}^{\infty}\psi(s,t;\mathrm{i}\lambda
)\,\psi(s,t;\mathrm{i}\mu)\,\mathrm{d}s,\qquad\lambda,\mu\geq0.
\]
Then $\mathbb{K}_{x,t}$ is trace class and positive, and
\[
q_{\sigma}(x,t)=q(x,t)-2\partial_{x}^{2}\log\det\!\bigl(I+\mathbb{K}%
_{x,t}\bigr)
\]
is again a step-type KdV solution with scattering data
\[
S_{q_{\sigma}}=\{\,R,\;\mathrm{d}\rho+\mathrm{d}\sigma\,\}.
\]

\end{theorem}

In the context of integrable systems, the \emph{binary Darboux transformation}
was introduced in \cite{Babichetal85} as a way to generate explicit solutions.
In our terminology it would correspond to a discrete finite measure $\sigma$.
However, in the spectral-theoretic setting it appeared even earlier as the
\emph{double commutation method} (see, e.g., Gesztesy--Teschl
\cite{GesztTeschl96} and the recent \cite{Ryb21} and the literature cited
therein). Theorem \ref{MainThm} represents its continuous counterpart. For
this reason we call it the \emph{continuous binary Darboux transformation},
since it performs the following transformation of scattering data:
\[
\{R,\mathrm{d}\rho\}\ \longrightarrow\ \{R,\mathrm{d}\rho+\mathrm{d}\sigma\}.
\]

Note that if the seed potential $q=0$, then Theorem~\ref{MainThm} clearly
reduces to Dyson's formula (\ref{Dyson}), which we have already discussed in
the context of soliton gases. There is, however, more to Theorem~\ref{MainThm}
than this. It readily offers a rigorous framework to construct deterministic
soliton gases on reflectionless (as well as arbitrary) step-like backgrounds
along the same lines as in Section \ref{sec:Gas}. To the best of our knowledge
this has not been rigorously developed elsewhere.

Another open problem comes from numerical experiments suggesting that
``injection'' of a soliton into a soliton condensate may locally in time and
space ``evaporate'' the latter, but this effect has yet to be described
mathematically. We believe that this phenomenon can be modeled within our
framework: a condensate background $\mathrm{d}\rho$ is perturbed by a narrowly
supported measure $\mathrm{d}\sigma$.

We conclude this section with some general remarks (see \cite{RybCBTD23}).

\begin{itemize}
\item There is no restriction on $\sigma$ beyond the integrability condition
$\int|\mathrm{d}\sigma|/k<\infty$, and therefore the negative spectrum may be
altered arbitrarily while the reflection coefficient remains unchanged.

\item The transformed potential $q_{\sigma}(x,t)$ is as smooth as the original
$q(x,t)$.

\item If $0\notin\operatorname{Supp}\sigma$, then $q_{\sigma}(x,t)-q(x,t)$
decays exponentially as $x\rightarrow+\infty$ for each fixed $t>0$.

\item If $\sigma(\{\kappa\})>0$ and $\kappa\in\operatorname{Supp}\rho$, then
$-\kappa^{2}$ becomes an \emph{embedded bound state} of $\mathbb{L}%
_{q_{\sigma}}$.

\item Depending on the sign of $\mathrm{d}\sigma$ we may add and/or remove
parts of the negative spectrum. Moreover the binary Darboux transformation is
invertible:
\[
\bigl(q_{\sigma}\bigr)_{-\sigma}=q.
\]

\item An analog of Theorem \ref{MainThm} can be stated for left scattering
data. Due to the directional asymmetry of KdV, however, some additional
restrictions must be imposed on the seed potential $q(x)$ (work in progress).
\end{itemize}

\section{Conclusion}

Back in 1971, Zakharov \cite{Zakharov(soliton gas)71} pioneered a statistical
description of multisoliton solutions (a \emph{rarefied soliton gas}), which
has attracted renewed attention in the present century after the introduction
of \emph{integrable turbulence} and a general framework for random solutions
of integrable PDEs in his influential paper \cite{Zakharov2009}. This
phenomenon was observed in shallow-water wind waves in Currituck Sound, NC
\cite{CostaOsbornePRL14} and was experimentally reproduced in a wave tank
\cite{RedoretalPRL19} and in optical fibers, drawing even greater interest
from a number of research groups (see, e.g.,
\cite{Condi22,ZakharovetalPhysD2016,El16,El21,Grava21,NabelikZakharov2019})
with different approaches.

\emph{Dense soliton gases and condensates}, particularly important from the
physical point of view, can be modeled as closures of pure soliton solutions
(cf.~\cite{ZakharovetalPhysD2016,El2016,Gesztesy-Duke92,ElKom05,El-et-al2011}%
). We mention in particular \cite{ZakharovetalPhysD2016}, where the
Zakharov--Manakov dressing method \cite{ZakhMan85} was used to produce
\emph{primitive potentials}, which are one-gap but neither periodic nor
decaying. Such solutions are parametrized by dressing functions $R_{1},R_{2}$,
and essentially only the case $R_{2}=0$ has been studied rigorously
\cite{Grava21} via Riemann--Hilbert techniques. For $R_{2}\neq0$ the only case
$R_{1}=R_{2}$ was considered in \cite{NabelikZakharov2019} (yielding an
elliptic one-gap potential if $R_{1}=R_{2}=1$), but the general case is still
out of reach. Note that the dressing method is not quite the inverse
scattering transform and cannot directly solve a Cauchy problem
\cite{Marchenko91}.

While seemingly unrelated at first glance, Theorem \ref{MainThm} places many
KdV soliton gas constructions into the context of the inverse scattering
method for the Cauchy problem and provides a rigorous framework in which
deterministic soliton gases on nontrivial backgrounds can be studied. In fact,
in the soliton gas community one is often interested in statistical quantities
(density of states, effective velocity, collision rate, etc.) for left
step-type KdV solutions of the form produced by Theorem \ref{MainThm} with
$q(x,t)=0$ (zero background) and specific absolutely continuous measures
$\mathrm{d}\sigma\geq0$ supported on intervals $[-a^{2},-b^{2}]$ with $b>0$.
The inclusion of $q(x,t)\neq0$ (nonzero backgrounds) and $b=0$ (small
solitons) into this picture is yet to be fully understood.

Another open problem comes from numerical simulations suggesting that
``injection'' of a soliton into a soliton condensate may locally in time and
space ``evaporate'' the condensate, but this effect has not been described
mathematically. Our framework suggests one way to model such scenarios: a
condensate background encoded by $\mathrm{d}\rho$ is perturbed by a small
measure $\mathrm{d}\sigma$ representing the injected soliton component.

We are yet to investigate these questions in detail, but at least Theorem
\ref{MainThm} alleviates concerns about the formal character of limiting
(scaling) arguments that are quite common in the physical literature on the
subject. It provides a robust operator-theoretic setting within which
deterministic soliton gases and condensates, including those on step-like
backgrounds, can be treated using the tools of modern spectral and scattering theory.

\section{Acknowledgment}

We are grateful to G. El and T. Bonnemain for numerous discussions on soliton gases.

\end{document}